\documentclass[prl,onecolumn,superscriptaddress,showpacs]{revtex4}

\usepackage{bm,amsmath,amssymb,graphicx}

\begin{document}

\title{Quantum-elastic bump on a surface}

\author{Victor Atanasov}
\affiliation{Faculty of Physics, Sofia University, 5 blvd J. Bourchier, Sofia 1164, Bulgaria}
\email{vatanaso@phys.uni-sofia.bg}
\homepage[web page ]{http://phys.uni-sofia.bg/~vatanaso/}

\author{Rossen Dandoloff}
\affiliation{Laboratoire de Physique Th\'{e}orique et
Mod\'{e}lisation , Universit\'{e} de Cergy-Pontoise, F-95302 Cergy-Pontoise, France}

\begin{abstract}
We use an exact solution of the  elastic membrane shape equation, representing the curvature,  which will serve as a quantum potential in the quantum mechanical two dimensional Schr\"odinger equation for a (quasi-) particle on the surface of the membrane. Surface curvature in the quasi one-dimensional case is related to an unexpected static formation: on one hand the elastic energy has a maximum where surface curvature has a maximum and on the other hand the concentration of the expectation value to find the (quasi-) particle is again where the elastic energy is concentrated, namely where surface curvature has a maximum. This represents a particular form of a conformon.
\end{abstract}


\maketitle

The interplay between geometry and quantum mechanics has intrigued physicists since the early stages of the development  of quantum mechanics\cite{1}. While teaching quantum mechanics, or especially applications of quantum mechanics to real physical systems, a question may arise in students' minds as to whether quantum mechanics can describe a curved physical system (especially in the domain of biophysics - polymers, membranes etc.). We give here a simple exactly solvable example and explain why curvature may be associated with attractive quantum geometrical potential. The explanation presupposes only the knowledge of Heisenberg's uncertainty principle. This may be taught to graduate students specializing in quantum physics.

An important role for that  interest has played the regained importance of geometry in physics since the appearance of  the special and the general theory of relativity\cite{MTW}. The works by Jensen and Kope and da Costa have given a boost of the quantum theory in curved spaces\cite{9}. Several works have been done in the areas of localization due to curvature of the space and the proper definition of linear momentum on curved surfaces\cite{AmJ}.

It has turned out that geometry can play the intermediary between two rather different domains in physical research, namely the classical theory of elasticity (thin rods whose statics and dynamics are governed by the Kirchhoff's equations) and quantum mechanics. It appears that thin rods and one dimensional quantum theory are governed by differential equations of the same form leading to the appearance of conformons: the localization of  elastic and electronic energy which may propagate in space without dissipation\cite{rossen}.

In fact many models for flexible polymers can be described by thin rods on the mechanical side. For example the wormlike chain (WLC) model \cite{3} with one elastic constant - the bending modulus. Another example is the static DNA
with its double-helix structure, which is described by the wormlike rod chain (WLRC) model \cite{4}
with an extra elastic constant - the twist rigidity. Considerable work has been
done on the equilibrium properties of these elastic models but their intrinsic dynamical
properties haven't been studied in depth\cite{rossen, victor}. The dynamical properties play an important role in the
energy and information flow along a bio-polymer, an issue of interdisciplinary interest.

The conformations present in these polymer models introduce a quasi-one-dimensional spaces where one should explore the quantum mechanics of various excitations. For example, the alpha-helical sections of the secondary structure of proteins have been proposed by Davydov as a medium for quantum energy transfer\cite{Davidov}.

Naturally, this line of investigation can be extended in two dimensions and encompass the quantum-elastic properties of membranes. In this paper we report yet another interesting example of the role of geometry in bringing together quantum theory and the theory of elasticity of membranes, leading to the appearance of 2d quantum-elastic bumps. The quantum-elastic bump is the region between two neighbouring domains to which we can assign elastic energies. Therefore the shape of the boundary is related to the minimization of the elastic energy of the boundary. The quantum aspect is associated with the presence of a (quasi-) particle between the domains exactly where the elastic energy is localized. 

Let us first note that the elastic energy density of a membrane is proportional to $H^2$ where $H$ is the mean curvature of the membrane.  The equilibrium shape equation for the membrane looks like a Schrodinger equation with the same potential (as the shape equation) proportional to $H^2-K$ where $K$ is the Gaussian curvature of the membrane.
 
We will present an exactly solvable case where $K=0$ and the equations are one dimensional nonlinear Schrodinger type of equations which have exact solutions. This solution will be used in the Schr\"odinger equation as quantum potential[2]

The quantum dynamics of a {\it nonrelativistic} (quasi-) particle constrained to an arbitrary orientable surface is well explored: the curvature of the surface induces an attractive (has a minimum where maximally curved) geometric potential due to R. C. T. da Costa \cite{9}
\begin{equation}
V_{G} = - \frac{\hbar^2}{2 m^{\ast}}\left( H^2-K \right) , 
\end{equation}
where $m^{\ast}$ is the effective mass of the particle, $\hbar$ is the
Planck's constant; $H=(\kappa_1 + \kappa_2)/2$ and $K=\kappa_1 \kappa_2$ are the Mean and the Gaussian curvature of the surface, respectively. Here $\kappa_1, \kappa_2$ are the two position-dependent principal curvatures of the surface\cite{DG}.

The concept of such an attractive quantum geometrical potential may be introduced to students by a simple quasi-one-dimensional system: a thin tube, whose axis is laying in the plane, and that is straight at its both ends and is bent in the middle.
The distance between two points laying on the axis of the tube from both sides of the bent $\Delta s$ is longer measured in the tube (measered along the axis) than the distance between them measured in the plane $\Delta x$. Following Heisenberg's uncertainty principle $\Delta s\Delta p_s \sim \hbar$ and $\Delta x\Delta p_x \sim \hbar,$ where $\sim$ means order of magnitude. Since $\Delta s >\Delta x$, that is the distance travelled along the arc-length is greater than the distance in flat coordinate space,  $\Delta p_s< \Delta p_x$ follows. The order of magnitude of the kinetic energy of a particle of mass $m$ in a volume of dimension $\Delta x$ is $\frac{\hbar^2}{m (\Delta x)^2 }$ (see \cite{LL}) and therefore $\frac{\Delta p_s^2}{2m}<\frac{\Delta p_x^2}{2m}$  i.e. the quantum particle has lower energy in the tube, near the bend, compared to a particle in the straight section of the tube.

The problem of a quantum particle constrained on a surface has been considered in \cite{9}.  
The main result is the separation of the three dimensional Schr\"odinger equation in a two dimensional surface part and an orthogonal to the surface part. The wave function $\xi$ is split into tangential $\xi_{\rm t}$ and normal $\xi_{\rm n}$ $(\xi=\xi_{\rm t}\xi_{\rm n})$.  The tangential and normal Schr\"odinger equations have the following form:

\begin{eqnarray}\label{Schrodinger for surface}
 -\frac{\hbar^2}{2 m^{\ast}}  \left[ \Delta_{S}  +
\left(  H^2 -  K         \right) \right] \chi_{\rm t} = i \hbar \frac{ \partial \chi_{\rm t}}{\partial t} && \\
\nonumber &&\\
\left[ - \frac{\hbar^2}{2 m^{\ast}} \frac{\partial^2 }{\partial
q_3^2}
 + V_{\lambda}(q_3) \right] \chi_{\rm n} = i \hbar \frac{ \partial \chi_{\rm n}}{\partial t} ,  &&
\end{eqnarray}

where $q_1, q_2$ are the coordinates on the surface given by a radius-vector $\vec{r}(q_1, q_2)$ with a metric tensor $$g_{ij}=\frac{\partial \vec{r}}{\partial q_i} . \frac{\partial \vec{r}}{\partial q_j}$$ and $V_\lambda$ is a very steep limiting potential that keeps the particle close to the surface. Hereafter $g={\rm det}\; g_{ij}. $ The coordinate $q_3$ measures distance in direction normal to the surface, that is  along the normal vector $\vec{n} = \frac{\partial \vec{r}}{\partial q_1} \times \frac{\partial \vec{r}}{\partial q_2} /  |\frac{\partial \vec{r}}{\partial q_1} \times \frac{\partial \vec{r}}{\partial q_2}| .$ The Laplace-Beltrami operator on the surface is given by\cite{DG}:
\begin{equation}\label{Laplace-Beltrami}
 \Delta_{S} =  \sum_{i,j=1}^{2} \frac{1}{\sqrt{g}} \frac{\partial}{\partial q_i}
\left( \sqrt{g} (g^{-1})_{ij} \frac{\partial }{\partial q_j}   \right).
\end{equation}
Equation (\ref{Schrodinger for surface}) will be used subsequently to show the correspondence between quantum and elastic properties.

Now we turn to the elastic energy of the membrane. The shape of membranes is due to the curvature of the membrane considered as a regular two-dimensional surface embedded in the Euclidean three-dimensional space. The elastic free energy of a piece of membrane is expressed in terms its curvature and surface energy $\lambda$ (akin to surface tension) which depends on the area of that surface. There are two types of curvature that characterize a surface: mean curvature $H$ and gaussian curvature $K.$  Only the mean curvature contributes to the elastic energy of the surface. On a surface with a given topology, integral over the surface of the gaussian curvature (via the Gauss-Bonnet theorem) is a constant, which depends on the genus (number of holes in the surface) of that surface.When varying the surface energy this constant does not contribute to the outcome. For closed surfaces the surface energy may also depend on the volume enclosed and the difference in pressure between the inside and the outside. The shape equation for the equilibrium conformation of membranes arises from a minimization argument.

The  functional for the shape energy due to Ou-Yang and Helfrich is \cite{Helfrich}
\begin{eqnarray}\label{elastic energy for surface}
 F = \frac12 k_c \oint \left( 2H - c_0  \right)^2 dS + \lambda \oint dS + \Delta p \int dV,
\end{eqnarray}
where $c_0$ is the spontaneous curvature of the membrane's surface, $k_c$ is the bending rigidity of the membrane, $\lambda$ is the membrane's tensile strength or surface tension, $\Delta p$ is the pressure difference between the upper and lower sides of the membrane.

Straight-forward but rather lengthy variational calculus  $\delta F =0$ yields the shape equation 
\begin{eqnarray}\label{general shape equation}
2 \lambda H - \Delta p & = &2 k_c  \Delta_{S} H +\\
\nonumber && k_c \left( 2 H^2 - 2 K - c_0 H \right) \left( 2H + c_0 \right).
\end{eqnarray}
It is beyond the scope of the present paper to reproduce this derivation but we refer the interested reader to \cite{Ou-Yang}.
In this equation $\Delta_{S}$ is the Laplace-Beltrami operator \cite{DG}. The shape equation is for the Mean curvature $H.$ Let us remind the reader that the functional $F$ that is varied depends on two independent variables. If the surface is given by $z(x,y),$ where $x$ and $y$ are independent variables, the corresponding Euler-Lagrange equation has the following form
\[
\frac{\partial F}{\partial z} - \frac{\partial}{\partial x} \frac{\partial F}{\partial \frac{\partial z}{\partial x} } - \frac{\partial}{\partial y} \frac{\partial F}{\partial \frac{\partial z}{\partial y} }=0.
\]

Suppose the membrane is open and immersed in homogeneous medium, then the pressure difference vanishes $\Delta p =0.$ In case of vanishing spontaneous curvature $c_0=0,$ which is only natural for symmetric membranes \cite{Helfrich}, the shape equation reduces to 

\begin{eqnarray}\label{shape equation}
\left[ \Delta_{S}  +2 \left(  H^2 -  K \right) \right] H (q_1, q_2) =  \epsilon^2 H, 
\end{eqnarray}
where $\epsilon^2 = { \lambda}/{ k_c } $

\begin{figure}[t]
\begin{center}
\includegraphics[scale=0.35]{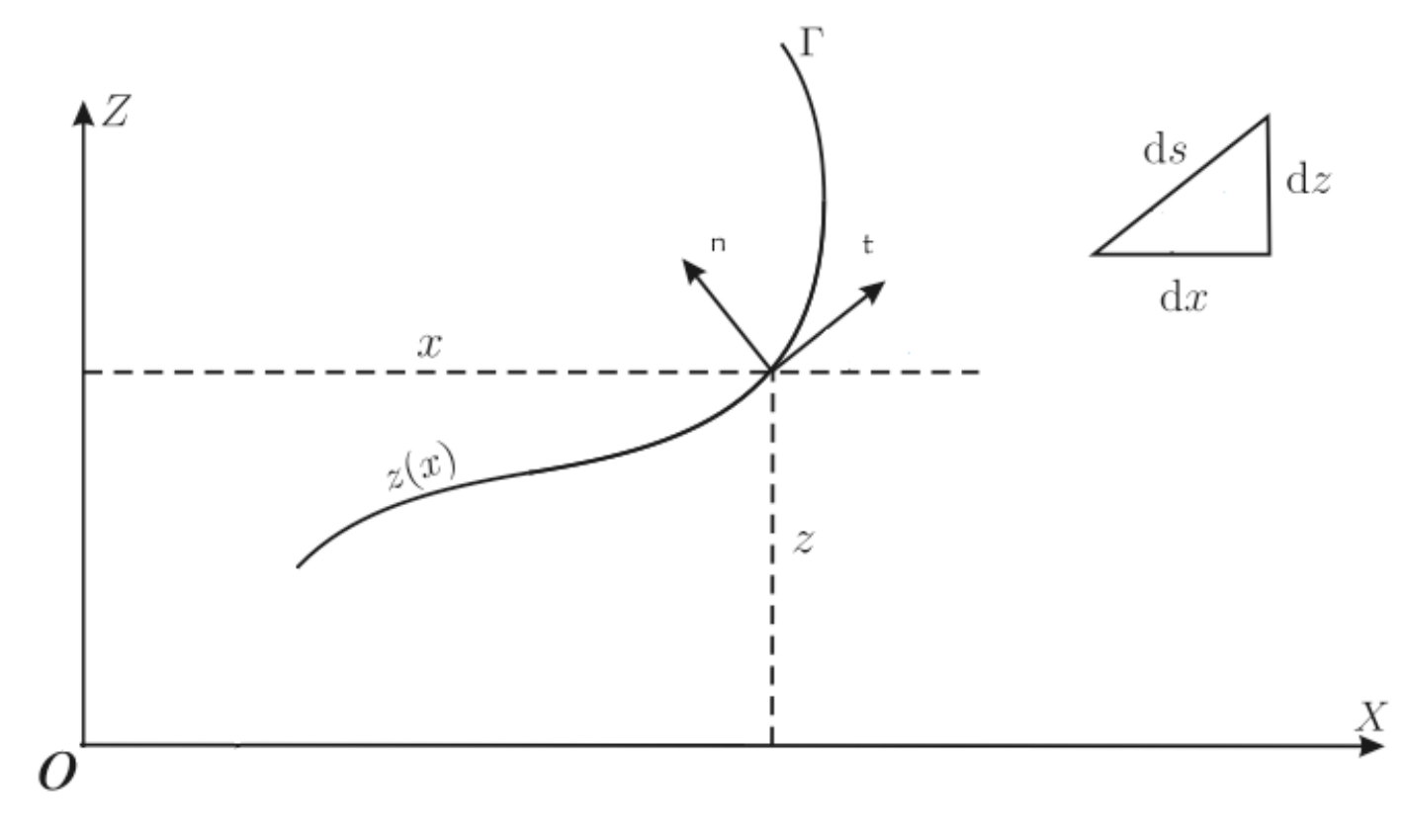}
\caption{\label{fig1} The profile curve $\Gamma.$ }
\end{center}
\end{figure}

We will present now a quasi one-dimensional exactly solvable case, which represents a translationally invariant solution to (\ref{shape equation}).
The translationally-invariant solutions are obtained by propagating along the y-axis the profile curve $\Gamma$ lying in the XOZ -plane. If $z(x)$ denotes the profile curve, then $s$ denotes the arclength along the curve. The following holds for the arclength: 
\begin{equation}\label{eq:arclength}
ds=\sqrt{ 1+ z'^2 } dx \quad z'=dz/dx.
\end{equation}
The translationally-invariant surfaces have vanishing Gaussian curvature $K=0$ due to $\kappa_2=0$ and $\kappa_1(x)=\kappa.$ Therefore $H=\kappa /2.$

One can represent the profile curve $\Gamma$ also by the graph $(x, z (x))$ of the function $z = z (x)$ (see Fig. \ref{fig1}). Employing standard calculation technique of \cite{DG}, the shape equation (\ref{shape equation}) reduces to the following nonlinear ordinary differential equation

\begin{eqnarray} \label{}
\frac{1}{\sqrt{1+z'^2}} \frac{d }{dx}\left( \frac{1}{\sqrt{1+z'^2}} \frac{d }{dx} \right) \kappa + \frac{1}{2} \kappa^3 = \epsilon^2 \kappa  
 \end{eqnarray}

which in terms of $s$ can be  further simplified

\begin{eqnarray} \label{eq:H}
\frac{d^2 \kappa}{ds^2} + \frac{1}{2} \kappa^3 = \epsilon^2 \kappa .
 \end{eqnarray}
Let us note that this equation coincides with the Kirchhoff's elastic rod equation, which has the same particular solution as the one discussed below\cite{rossen, goriely}. The solution to the shape equation (\ref{eq:H}) for a translationally-invariant surface  is soliton-like:

\begin{equation}\label{eq:soliton}
\kappa(s) = 2 \epsilon \; {\rm sech}(\epsilon s).
\end{equation}

Finding solutions to (\ref{eq:H}) having the property of non-constant mean curvature $H(s)$ is an intriguing problem leading to major consequences such as localization of elastic energy. 

Inserting $\chi_{\rm t} = \psi (q_1, q_2) e^{i E/ \hbar t}$ in to (\ref{Schrodinger for surface}) leads to the stationary Schr\"odinger equation on the surface

\begin{eqnarray}\label{stationary Schrodinger for surface}
  \left[ \Delta_{S}  +
\left(  H^2 -  K         \right) \right] \psi (q_1, q_2) = \varepsilon^2  \psi
\end{eqnarray}
where $\varepsilon^2 = { 2 m^{\ast} E }/{\hbar^2} $

For the quasi one dimensional case we will demonstrate that localisation of electronic energy is possible as well. Let us turn to the Schr\"odinger equation (\ref{stationary Schrodinger for surface}) on the surface

\begin{eqnarray} \label{}
\frac{1}{\sqrt{1+z'^2}} \frac{d }{dx}\left( \frac{1}{\sqrt{1+z'^2}} \frac{d }{dx} \right) \psi + \frac{1}{4} \kappa^2 \psi = \varepsilon^2 \psi.  
 \end{eqnarray}

which in terms of $s_1=s/\sqrt{2},$ that is the  rescaled arclength along the curve as defined in (\ref{eq:arclength}), becomes:

\begin{eqnarray} \label{}
\frac{d^2 \psi}{ds_1^2} + \frac{\left[\kappa(s_1)\right]^2}{2}  \psi = 2 \varepsilon^2 \psi.  
\end{eqnarray}

where $k(s_1)$ is given in eq(12).

\begin{figure}[b]
\begin{center}
\includegraphics[scale=0.5]{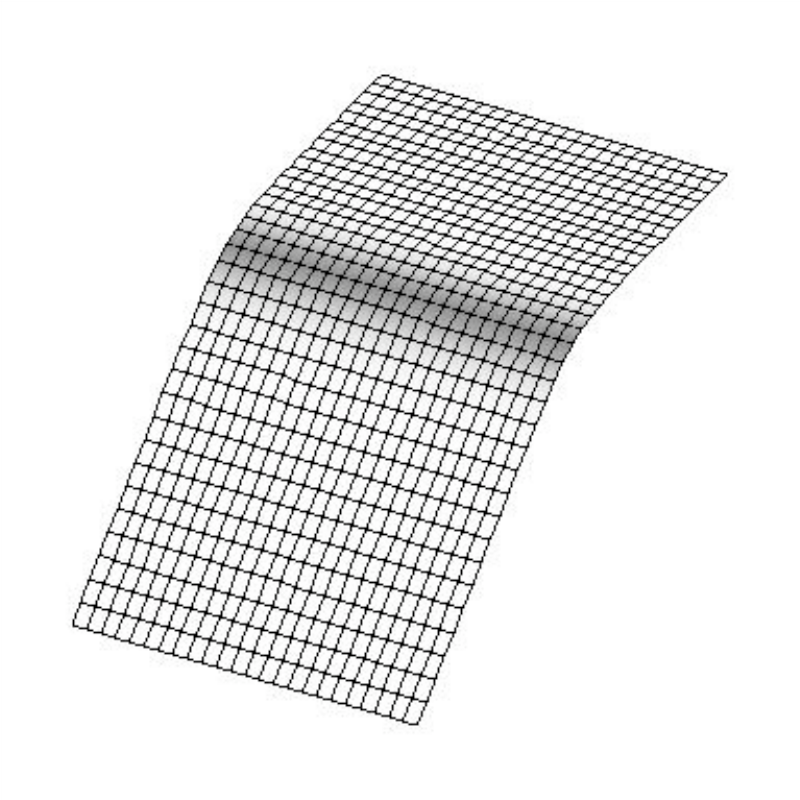}
\caption{\label{fig2} The profile of the surface, where the maximum probability density on the surface is represented by the dark strip.  The boundary conditions are such that the curvature and the wave function at infinity along the arclength go to zero.}
\end{center}
\end{figure}

Looking for a solution for the wave function of the same form as the surface curvature (\ref{eq:soliton}) we see that 

\begin{equation}\label{}
\psi =\kappa(s_1)= 2 \epsilon {\rm sech}(\epsilon s_1),
\end{equation}
satisfies the Schr\"odinger equation, where $\epsilon^2=2 \varepsilon^2$. The visual representation of the soliton-like surface and the corresponding probability density $|\psi|^2$ is given in Fig.\ref{fig2}. Note that the position of the maximum of the elastic energy and probability density coincide. This points to an interesting occurance of localization of the probability density to find  a (quasi-) particle  where the surface is maximally curved. This represents a conformon. Conformon represents a localization of at least two types of energies, for example: electronic and elastic, magnetic and elastic, ect. Indeed, this conformon separates two domains of the surface where the curvature is vanishing. In this case it appears as a quantum-elastic bump along one direction on the surface.

In conclusion, we state the main observation in the paper: the shape equation for an elastic open membrane is similar to the Schr\"odinger equation on the surface. The main consequence in the quasi one-dimensional case (where one of the principal curvatures is zero) we considered is the concentration of the probability density for a (quasi-) particle on the surface where its curvature has a maximum (the elastic energy has a local maximum too). On the other hand the curvature represents the quantum potential in the Schr\"odinger equation which transforms into a nonlinear Schr\"odinger equation and the solution for the wave function is again $ k(s_1)$. This mechanism represents a spacial case of a conformon. A similar mechanism has been proposed in [4]. The proposed model  represents an interest in bio and membrane physics as well as microelectronics involving graphene and graphene oxide .

\end{document}